\documentclass[letter]{aa} % for the letters 
\usepackage{graphicx}
\usepackage{txfonts}
\usepackage{natbib}
\bibpunct{(}{)}{;}{a}{}{,} % to follow the A&A style%

\begin{document}
\title{Effects of rotation and magnetic fields on the lithium abundance and asteroseismic properties of exoplanet-host stars}
\titlerunning{Rotation, magnetic fields and the lithium abundances of exoplanet-host stars}

\author{P. Eggenberger \and A. Maeder \and G. Meynet}

   \offprints{P. Eggenberger}

\institute{Observatoire de Gen\`eve, Universit\'e de Gen\`eve, 51 Ch. des Maillettes, CH-1290 Sauverny, Suisse\\ 
	\email{[patrick.eggenberger;andre.maeder;georges.meynet]@unige.ch}
             }

   \date{Received; accepted}

% \abstract{}{}{}{}{} 
% 5 {} token are mandatory
 
  \abstract
   {Recent spectroscopic observations indicate lower lithium abundances of exoplanet-host stars 
compared to solar-type stars without detected exoplanets. Moreover, studies based on the distribution
of rotational periods of solar-type stars suggest a possible link between lithium depletion and the rotational
history of exoplanet-host stars.}
   {The effects of rotation and magnetic fields on the surface abundances of solar-type stars are studied in order
to investigate whether the reported difference in lithium content of exoplanet-host stars can be related to their
rotational history. Moreover, the asteroseismic properties predicted for stars with and without exoplanets are compared
to determine how such a scenario, which relates the lithium abundances and the rotational history of the star, can be
further challenged by observations of solar-like oscillations.}
   {Based on observations of rotational periods of solar-type stars, slow rotators on the zero age main sequence (ZAMS) are modelled with a comprehensive treatment of only the shellular rotation, while fast rotators are modelled
including both shellular rotation and magnetic fields. 
Assuming a possible link between low rotation rates on the ZAMS and the presence of planets as a result of a longer disc-locking phase during the pre-main sequence (PMS), we compare the surface abundances and asteroseismic properties
of slow and fast rotating models, which correspond to exoplanet-host stars and stars without detected planets, respectively.}
   {We confirm previous suggestions that the difference in the lithium content of stars with and without detected planets
can be related to their different rotational history. The larger efficiency of rotational mixing 
predicted in exoplanet-host stars explains their lithium depletion and also leads to changes 
in the structure and chemical composition of the central stellar layers. Asteroseismic observations can reveal
these changes and can help us distinguish between different possible explanations for the lower lithium content of exoplanet-host stars.}
   {}

   \keywords{stars: solar-type -- stars: abundances -- stars: rotation -- stars: magnetic field -- stars: oscillation}

   \maketitle
%
%________________________________________________________________

\section{Introduction}

Today we know of more than 400 exoplanets, which renders the understanding
of planetary systems formation and evolution a key question.
Into this context belongs the growing interest in the comparative study of the properties of exoplanet-host stars 
to similar stars without detected planets. \cite{isr09} report an important result concerning the lithium abundances of stars with and without exoplanets.
A significant lithium depletion is indeed found for exoplanet-host stars compared to solar-type stars without
detected exoplanets in the effective temperature range 5600--5900 K.
Recalling that light elements are valuable tracers of transport mechanisms at work
in the external stellar layers, these observations immediately raise the question 
which physical process is responsible for these differences in the lithium content of
stars with and without detected exoplanets.    

Rotation is one of the key processes 
that change the internal structure and global properties of stellar models 
with a peculiarly strong impact on the lithium surface-abundances \cite[see e.g.][]{pin10}. 
The interesting connection between the lithium depletion, the rotational
history of solar-type stars and the presence of exoplanets has been recently discussed by \cite{bou08}.
The first important observational fact in this context is that the rotational periods 
of young solar-type stars suggest that slow rotators develop a high degree of differential rotation between the radiative core 
and the convective envelope, while solid-body rotation is favoured for fast rotators \citep[see e.g.][]{irw07,bou08,denetal10}. 
The rotation of the star on the ZAMS basically depends on its initial velocity 
and on the disc lifetime during the pre-main sequence. A longer disc lifetime enables the star to lose 
a larger amount of angular momentum during the PMS, which leads to a lower rotation rate on the ZAMS. 
Because longer disc lifetimes may favour the formation and migration of giant exoplanets, 
low rotation rates on the ZAMS may be expected for exoplanet-host stars, as suggested by \cite{bou08}.
Finally, the differential rotation 
between the radiative core and the convective envelope found for slow rotators also suggests a 
lower lithium abundance for exoplanet-host stars, which agrees well with
the observations of lithium abundances reported by \cite{isr09}. A longer disc lifetime may therefore lead 
simultaneously to a lower lithium abundance (owing to the lower rotation rate on the
ZAMS) and a higher probability to detect giant exoplanets. 

This scenario proposed by \cite{bou08} 
is based on rotational models that do not include a detailed physical description of
the transport processes that are at work in stellar interiors. Moreover, the lithium depletion is only qualitatively 
discussed by \cite{bou08}. We here study first the lithium abundances expected for stars with and without exoplanets 
by computing stellar models that include a comprehensive treatment of shellular rotation
and the effects of internal magnetic fields as prescribed by the Tayler-Spruit dynamo \citep{spr02}. 
The asteroseismic properties of stars with and without exoplanets are then investigated 
to determine new observational diagnostics that are able to test such a scenario, which relates
the lithium abundance and presence of exoplanets to the rotational history of the star.

The effects of rotational mixing and internal magnetic fields on the lithium abundance of
solar-type stars with and without exoplanets are studied in Sect.~2. The asteroseismic properties of these stars are discussed in Sect.~3, while 
we give the conclusion in Sect.~4.

\section{Effects of rotation and magnetic fields on the lithium abundances of exoplanet-host stars}
\label{sec_li}

The stellar evolution code used for these computations is the Geneva code,
which includes shellular rotation and the Tayler-Spruit dynamo as described in \cite{egg08}. 
In addition to shellular rotation and magnetic fields, atomic diffusion is included and the diffusion coefficients 
are computed according to the prescription by
\cite{paq86}.

\begin{figure}[htb!]
 \resizebox{\hsize}{!}{\includegraphics{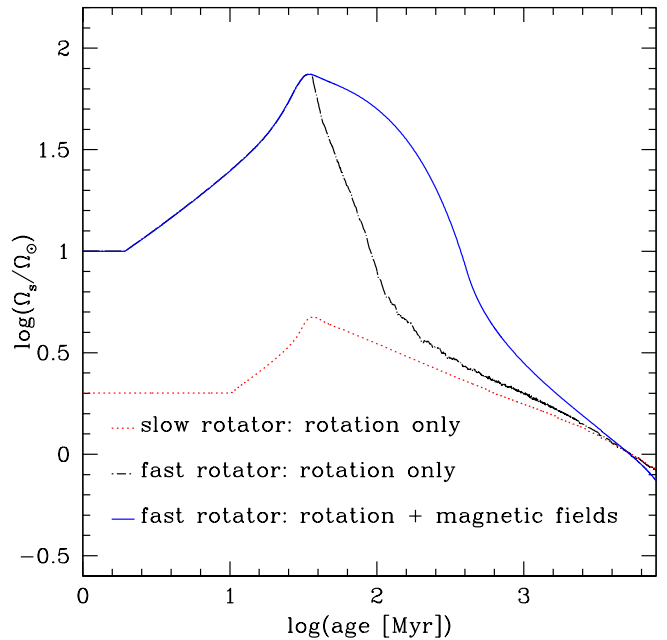}}
  \caption{Variation with time of the surface angular velocity of 1\,M$_\odot$ models for fast and slow rotators on the ZAMS. The model for slow
rotators (dotted line) is computed with rotation only, while the model for fast rotators (continuous line) includes both rotation and magnetic fields. The evolution of a model for fast rotators computed with rotation only is shown by the dot-dashed curve.}
  \label{omegast}
\end{figure}

As mentioned in the introduction, the observation of rotational periods of young solar-type stars 
suggests that slow rotators develop a high degree of differential rotation between the radiative core 
and the convective envelope, while solid-body rotation is favoured for fast rotators. 
Because shellular rotation alone produces an internal coupling that is insufficient to ensure
solid-body rotation \citep{pin89,cha95}, this suggests that another effect intervenes for fast rotators.
Two main mechanisms have been proposed to efficiently extract angular momentum from the central
core of a solar-type star: magnetic fields \citep[e.g.][]{mes87,cha93,egg05_mag,den10} 
and internal gravity waves \citep[e.g.][]{scha93,kum97,zah97,cha05,tal05}.  
We recall here that the theoretical prescription for a dynamo operating in the radiative zone of a star \citep[see e.g.][]{den07} as well
as its real existence \citep{zah07,gel08} is
still a matter of debate. It is beyond the scope of the present paper 
to study in detail the theoretical description of such a dynamo and in particular the physical conditions needed for this mechanism to operate. We simply note that observations of rotational rates of solar-type stars indicate that slow rotators on the ZAMS seem to exhibit internal differential rotation, while a strong internal coupling is favoured for fast rotators on the ZAMS.  Consequently, slow rotators are modelled in the present study with shellular rotation only, while
fast rotators are modelled with both rotation and the Tayler-Spruit dynamo to produce an 
internal coupling that is sufficient to ensure solid-body rotation. Note that the inclusion of both rotation and magnetic fields is assumed for fast rotators on the ZAMS during the whole evolution. The variation with time of the surface velocity of these models for fast and slow rotators 
on the ZAMS is shown in Fig.~\ref{omegast}, together with the evolution of the surface velocity of a model for fast rotators including shellular rotation only. 
Figure~\ref{omegast} clearly shows that shellular rotation alone does not 
couple the core and envelope even in the presence of rapid rotation.
This illustrates the need to introduce an efficient mechanism for the transport of angular momentum for fast rotators, because the rapid decrease of the surface angular velocity observed for this model cannot correctly reproduce the spin-down of fast rotators in open clusters \citep[e.g.][]{den10}.

\begin{figure}[htb!]
 \resizebox{\hsize}{!}{\includegraphics{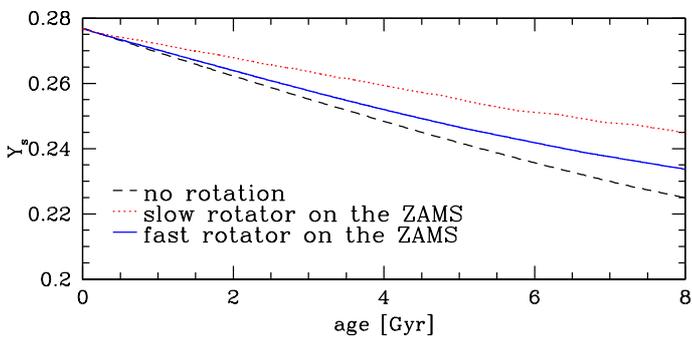}}
  \caption{Surface abundance of helium $Y_{\rm s}$ during the evolution 
of 1\,M$_\odot$ models. The model of a slow rotator on the ZAMS is computed with rotation only, while
the model of a fast rotator on the ZAMS is computed with both rotation and magnetic fields.}
  \label{yst_exo}
\end{figure}

We compute the evolution of 1\,M$_{\odot}$ models with a solar chemical composition 
as given by \cite{gre93} and a solar calibrated value for the mixing-length parameter. 
We adopt the braking law of \cite{kri97} with $\Omega_{\rm sat}=8$\,$\Omega_\odot$ to reproduce the magnetic braking undergone by solar-type stars. The braking constant is then calibrated so that rotating models
reproduce the solar surface rotational velocity after $4.57$\,Gyr.
Note that the transport of chemical elements is
sensitive to the assumed magnetic braking prescription. While the exact values of light element abundances at the stellar surface
can be somewhat changed by using different prescriptions for
this braking law, the results remains qualitatively the same.
One model is first computed without 
the inclusion of rotation or magnetic fields. A model for slow rotators is computed with shellular rotation
only and an initial angular velocity of 2\,$\Omega_\odot$, which remains coupled to its disk for 10\,Myr. 
The corresponding model for fast rotators
is computed with both rotation and the Tayler-Spruit dynamo and an initial angular velocity of 10\,$\Omega_\odot$, which 
remains coupled to its disk for 2\,Myr. The evolution of the surface velocity of these models is shown in Fig.~\ref{omegast}.
To study the surface abundances of slowly and fast rotating solar-type stars,
the variation of the helium surface abundance $Y_{\rm s}$ during the main sequence is plotted 
in Fig.~\ref{yst_exo} for these models. Owing to atomic diffusion, the helium abundance at the surface of the non-rotating
model (dashed line) steadily decreases during the main-sequence evolution.
Rotational mixing counteracts the effects of atomic diffusion in the external layers of the star. As a result, the slowly rotating
model (dotted line) exhibits higher values of the surface helium abundance than the non-rotating model
(dashed line). The fast rotating model (continuous line) exhibits
slightly higher values of the helium surface abundance than the non-rotating model, but lower values than
the model for the slow rotator. This is because the efficiency of rotational mixing in the external
layers of solar-type stars is strongly reduced when the effects of the Tayler-Spruit dynamo are taken into account.
This is a direct consequence of the near solid-body rotation of models that include both
rotation and magnetic fields. The slowly rotating model computed with shellular rotation only 
is characterized by a higher degree of differential rotation between the radiative core 
and the convective envelope than the fast rotating model that includes both rotation and magnetic fields. 
There is however a slight increase of the transport of chemicals by the meridional circulation 
for models with magnetic fields, which is also due to the near solid-body rotation, 
because uniform rotation creates a strong breakdown of the local radiative equilibrium. For solar-type
stars, this increase is much smaller than the strong decrease of the shear turbulent-mixing though, which leads
to a net decrease of the efficiency of overall rotational mixing for slow rotators compared to fast rotators.

\begin{figure}[htb!]
 \resizebox{\hsize}{!}{\includegraphics{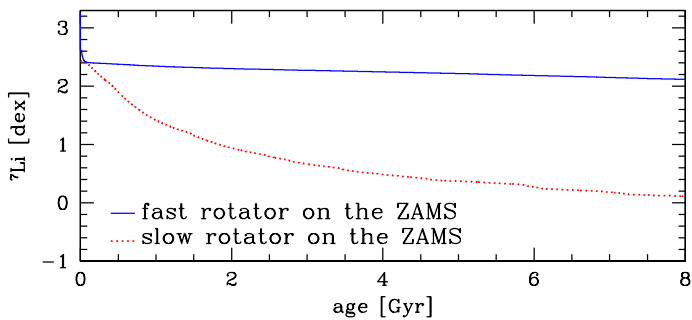}}
  \caption{Same as Fig.~\ref{yst_exo} for the surface abundance of lithium.}
  \label{ab_lithium}
\end{figure}

Figure~\ref{ab_lithium} compares the evolution of the 
surface lithium abundance of the models for slow and fast rotators starting with a lithium abundance 
$A({\rm Li})=\log [N({\rm Li})/N(H)]+12=3.26$  \citep[][]{asp09}. 
Owing to the strong decrease of the efficiency of shear turbulent mixing when magnetic fields are taken
into account, the model for fast rotators (continuous line) only shows a very limited decrease of its lithium content
during the main sequence compared to the model for slow rotators (dotted line). Figure~\ref{ab_lithium} clearly
shows that slow rotators are more lithium-depleted than fast rotators.

Low rotation rates on the ZAMS are assumed for exoplanet-host stars, 
because longer disc lifetimes may favour the formation and migration of giant exoplanets \citep[e.g.][]{bou08}. Consequently,
exoplanet-host stars are predicted to be more lithium-depleted than solar-type stars without
detected exoplanets. By computing models of solar-type stars with shellular rotation and 
the Tayler-Spruit dyamo, we thus confirm the suggestion of \cite{bou08} that exoplanet-host stars are more
lithium-depleted because of the higher efficiency of rotational mixing in these stars, which are slow rotators on the ZAMS.
In this scenario, a relationship between lithium abundance and surface rotational velocity is thus also
predicted for young solar-type stars, which agrees well with observations in the Pleiades 
of higher lithium abundances for fast rotators than for slow rotators \citep{sod93}. This correlation between
the lithium content and the surface velocity is of course only predicted for solar-type stars at the beginning of
the main sequence, because the surface velocity of fast and slow rotators on the ZAMS rapidly converges to the same value
due to the magnetic braking undergone by solar-type stars (see Fig.~\ref{omegast}).

\section{Asteroseismic properties of exoplanet-host stars}

The higher efficiency of rotational mixing predicted in exoplanet-host stars than in stars without planets
can explain the different lithium abundances of these stars, 
but this also leads to changes in the structure and chemical composition of the
central layers. Because the effects of rotational mixing on the properties of the central layers 
of a solar-type star can be revealed by asteroseismic observations, 
we are now investigating the asteroseismic properties of stars with and without planets. 

We compare models of stars with and without planets that share
the same luminosity, effective temperature, and surface metallicity. 
For this purpose, we computed another model for fast rotators on the ZAMS.
The initial chemical composition and mixing-length parameter of this 1\,M$_{\odot}$ model
with both rotation and the Tayler-Spruit dyamo is calibrated to simultaneously reproduce  
the surface metallicity and location in the HR diagram of the slowly rotating model 
after 6\,Gyr. In addition to decreasing the Li abundance, the higher efficiency of
rotational mixing for the model of an exoplanet-host star also changes the surface abundances of $^3$He, Be, and B
by 0.3, 0.9 and 0.15\,dex, respectively. Only negligible changes of the $^{12}$C/$^{13}$C ratio and of the main-sequence lifetime 
are found for models of exoplanet-host stars.

\begin{figure}[htb!]
 \resizebox{\hsize}{!}{\includegraphics{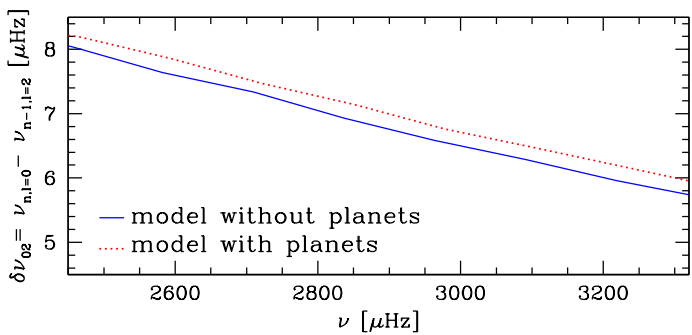}}
  \caption{Variation with frequency of the small separation between $\ell=0$ and $\ell=2$ modes 
for 1\,M$_\odot$ models with the same metallicity and location in the HR diagram.}
  \label{pt_magn_t6}
\end{figure}

The theoretical low-$\ell$ frequencies of both models are then computed with the Aarhus adiabatic 
pulsation code \citep{chr08}.
The slowly rotating model which corresponds to 
a model for an exoplanet-host star is characterized by a mean large frequency separation of 127.3\,$\mu$Hz after 6\,Gyr, 
while the fast rotator on the ZAMS has a mean large frequency separation of 127.4\,$\mu$Hz.
The mean large frequency separation is mainly sensitive to the mean density of the star.
For this comparison, both models have the same mass and radius, which leads to similar values of the large frequency separation.
The small frequency separation is very sensitive 
to the structure of the stellar core and is mainly proportional to the hydrogen content of the central layers.
The variation with frequency of the small separation $\delta \nu_{02}$ 
between oscillation modes with $\ell=0$ and $\ell=2$ is compared for both models in Fig.~\ref{pt_magn_t6}.
This figure shows that the model for an exoplanet-host star exhibits higher values of the small separation
than the corresponding model without planets. This is due to the change by rotational mixing of the chemical
profiles in the stellar interior of the exoplanet-host star model and in particular to the increase of the 
central abundance of hydrogen. This illustrates that the 
strong reduction of the efficiency of rotational mixing when the effects of internal magnetic
fields are taken into account is not limited to the external stellar layers, but is
also observed in the deep interior of solar-type stars. The transport of fresh hydrogen fuel in the
central stellar core is more efficient for the slow rotating model on the ZAMS computed with rotation only 
than for the initially fast rotating model with both rotation and magnetic fields, which leads to higher values
of the small frequency separations for models of stars with planets than for stars without planets. Owing to the
low initial velocity of the exoplanet-host star model, the increase
of the small frequency separation is somewhat limited. Figure~\ref{pt_magn_t6} indeed shows
that a difference of about 0.2\,$\mu$Hz is typically predicted between the mean small frequency separation of the
model with and without exoplanets at an age of 6\,Gyr. This small frequency difference requires high-precision asteroseismic observations 
during a long time to reach the required frequency resolution. In this context, asteroseismic 
observations coming from space missions will be particularly valuable, because first results of the {\it Kepler} mission
indicate that a typical precision of 0.2\,$\mu$Hz on the mean small frequency separation of solar-type stars is
already reached after the first 33.5 days of science operations \citep{cha10}, and that very high precisions on the
frequency separations will then be obtained with the continuously increasing total observing time.

\section{Conclusion}

The effects of rotation and magnetic fields on the surface abundances of solar-type stars have been 
first studied in the framework of a scenario suggested by \cite{bou08}, which relates the enhanced lithium depletion in exoplanet-host
stars to their rotational history. This scenario is based on the observations of rotational periods 
of solar-type stars, which suggest differential rotation between the radiative core and the convective envelope 
for slow rotators on the ZAMS and solid-body rotation for fast rotators. These observations thus
indicate that a very efficient process for the transport of angular momentum is at work in fast rotators. 
Slow rotators were then modelled with shellular rotation only, while
fast rotators were modelled with both rotation and the Tayler-Spruit dynamo to 
ensure solid-body rotation. It is thus important to note that the different mixing 
properties observed in this work between slow and fast rotators mainly come from the inclusion of internal
magnetic fields for fast rotators and not from an intrinsic property of rotation observed in slow and fast rotating
models computed with exactly the same input physics.
As a result of the strong decrease of the efficiency of shear 
turbulent mixing when magnetic fields are taken into account, slow rotators are found to be more
lithium depleted than fast rotators. Assuming that the presence of giant exoplanets is favoured
for stars with slow rotation rates on the ZAMS due to a longer disc lifetime during the PMS evolution,
lower lithium surface abundances are found for models of exoplanet-host stars compared 
to solar-type stars without planets, which agrees well with the spectroscopic observations of \cite{isr09}.
These results show that the lithium depletion observed at the surface of exoplanet-host stars 
is possibly related to the rotational history of solar-type stars, but this is of course not 
the only possible explanation. An increase of the mixing efficiency triggered by angular momentum 
transfer due to planetary migration \citep{cas09} or thermohaline convection below the convective zone
associated to the accretion of planetary material onto the star in its early phases \citep{the10} 
can also provide valid explanations for the observed dispersion of lithium abundances 
in solar-type stars with and without planets. To distinguish between these explanations,
we find that in the scenario studied here relating the lithium abundance of the 
star to its rotational history, the higher efficiency of rotational mixing in exoplanet-host stars not only 
changes the surface stellar abundances, but also leads to changes in the structure 
and chemical composition of the central layers, which can be revealed by asteroseismic observations. In particular,
higher values of the small frequency separations are predicted for exoplanet-host stars than
for stars without planets. It will be particularly interesting to investigate this
point in the light of new asteroseismic observations coming from space missions that are dedicated simultaneously to
the search of exoplanets and asteroseismology like CoRoT and {\it Kepler}.

\begin{acknowledgements}
We thank J. Christensen--Dalsgaard for providing us with the Aarhus adiabatic pulsation code.
This work was supported by the Swiss National Science Foundation.
\end{acknowledgements}

\bibliographystyle{aa} % style aa.bst
\bibliography{biblio} % your references Yourfile.bib

\end{document}